\documentclass[12pt]{article}

\begin{document}
\begin{center}
{\bf Pair Production and Vacuum Polarization of Arbitrary Spin
Particles with EDM and AMM}\\
\vspace{5mm}
 S.I. Kruglov \\
\vspace{5mm}
\textit{International Educational Centre, 2727 Steeles Ave. W, \# 202, \\
Toronto, Ontario, Canada M3J 3G9}
\end{center}

\begin{abstract}
The exact solutions of the wave equation for arbitrary spin particles with
electric dipole and magnetic moments in the constant and uniform
electromagnetic field were found. The differential probability of pair
production of particles by an external electromagnetic field has been
calculated on the basis of the exact solutions. We have also estimated the
imaginary part of the constant and uniform electromagnetic field. The
nonlinear corrections to the Maxwell Lagrangian have been calculated taking
into account the vacuum polarization of arbitrary spin particles. The role
of electric dipole and magnetic moments of arbitrary spin particles in
instability of the vacuum is discussed.
\end{abstract}

\section{Introduction}

The electric dipole moment (EDM) of particles violates the $CP$ invariance,
the time-reversal ($T$) symmetry [1] and may be induced by the $\vartheta $
-term of the quantum chromodynamics (QCD) vacuum. In particular, the neutron
and vector mesons may possess the EDM due to the $\vartheta $-term which
violates $P$ and $CP$ symmetries and gives $CP$-odd electromagnetic
observable [2-4]. It should be noted that the $\vartheta $-term is important
for the solution of the $U(1)_A$ problem in strong interactions. The
presence of the $\vartheta $-term creates the ``strong $CP$ problem" which
can be solved by introducing axions - scalar fields with the small mass. The
experimental upper bound for the neutron leads to the experimental
constraint of the EDM of the W-boson: $d_W\leq 10^{-19}e$ cm [5]. In the
standard model (SM) the predicted EDM of the W-boson is $d_W\simeq 10^{-29}e$
cm [4]. The EDM of particles can be induced by the Higgs-boson exchange [6].
At the same time the EDM of the W-boson may contribute to the EDM of
fermions (for example electrons). The experimental measurements of the EDM
of elementary particles is important to verify the status of the SM.

The $CP$ - violation observed in the decays of mesons is a fundamental
phenomenon and remains mysterious. In the SM the $CP$ - violating
interactions can be introduced by the Kobayashi-Maskawa matrix, and the
predicted EDM's of elementary particles are extremely small. Some SUSY and
multi-Higgs models may predict much stronger $CP$ - violating effects [7].

The properties of hadrons can be investigated in the framework of the
renormalizable theory of strong interactions of quarks and gluons - QCD.
However, the main tool in QCD is the perturbation theory in small coupling
constant $\alpha _s$ which works only in the ultraviolet region at high
energy. The characteristics of hadrons are described in the infrared region
of QCD at low energy where the nonperturbative effects of chiral symmetry
breaking and confinement of quarks take place. The nonperturbative methods
in QCD have not developed yet, and some models of hadrons are used. It
should be noted that in the framework of the QCD string theory [8] mesons
and baryons possess the EDM [9]. It is very important to study various
processes involving arbitrary spin particles with the EDM and anomalous
magnetic moments (AMM) in view of the great interest to physics in framework
and beyond the SM.

Here we proceed from the second order relativistic wave equation for
arbitrary spin ($s$) particles with the EDM and AMM on the basis of the
Lorentz representation $(0,s)\oplus (s,0)$ of the wavefunction. In the case
of ``normal'' magnetic moment and the absence of the EDM, such a scheme was
introduced in [10]. The wavefunction in this approach has the minimal number
$2(2s+1)$ of components, and particles propagate causally in external
electromagnetic fields.

It is interesting to study the pair production probability and the vacuum
polarization of particles because there is the vacuum instability of
particles in a magnetic field [11-13]. In particular, the vacuum of vector
particles is non-stable in a magnetic field as there is a contribution to
the negative part of the Callan-Symanzik $\beta $-function, and the vacuum
is reconstructed in a magnetic field.

The problem of the pair production and vacuum polarization of
vector particles with gyromagnetic ratio $2$ was investigated in
[14-17]. This case corresponds to the renormalizable SM where
there is a certain symmetry of the vector electromagnetic
vertices. The gyromagnetic ratio for $W$-bosons $ g=1+\kappa $,
and the AMM $\kappa =1$. However, for vector particles in the
framework of the Proca Lagrangian, the gyromagnetic ratio $g=1$,
and the AMM $\kappa =0$. The pair production probability of
arbitrary spin particles with the gyromagnetic ratio $g$ was
studied in [11] on the basis of the semiclassical imaginary-time
method. However, this method is valid only for weak
electromagnetic fields. In [18] we found the pair production
probability for higher spin particles with the gyromagnetic ratio
$g$ for arbitrary external fields. Here we generalize this result
on the case of the arbitrary spin particles with the EDM, and
study also the vacuum polarization.

The purpose of this paper is to find exact solutions to equations for
arbitrary spin particles which possess the EDM and AMM in uniform and
constant electromagnetic field, and to use these solutions for investigating
the most interesting and important vacuum quantum effects - pair production
and vacuum polarization. The found exact solutions of the equations can also
be used for the other calculations of the electromagnetic processes with the
presence of arbitrary spin particles with the EDM and AMM.

The paper is organized as follows. In section 2 we find exact solutions of
the wave equation for arbitrary spin particles with the EDM and AMM in
external constant and uniform electromagnetic fields. The differential pair
production probability of particles and the imaginary part of the effective
Lagrangian for electromagnetic fields are calculated on the basis of exact
solutions in section 3. In section 4 the nonlinear corrections to Maxwell's
Lagrangian caused by the vacuum polarization of arbitrary spin particles
with the EDM and AMM are evaluated using the Schwinger method. Section 5
contains the conclusion.

The system of units $\hbar =c=1$, $\alpha =e^2/4\pi =1/137$, $e>0$ is used.

\section{Arbitrary spin particles with EDM and AMM in
electromagnetic fields}

Let us consider the theory of arbitrary spin particles in external
electromagnetic fields. We imply that the wavefunction for massive particles
is transformed under the $(s,0)\oplus (0,s)$ - representation of the Lorentz
group, where $s$ is the spin of particles. The wavefunction in this
representation possesses $2(2s+1)$ components. The Lorentz group
representations $(s,0)$, $(0,s)$ are parity conjugated and have $(2s+1)$
components each in accordance with the number of spin projections $s_z=\pm
s,\pm (s-1),...$ .In the case of the spin-$1/2$ particles the $(1/2,0)\oplus
(0,1/2)$ - representation corresponds to Dirac bispinors. For vector
particles we arrive at the representation $(1,0)\oplus (0,1)$ of the second
rank antisymmetric tensor. In this case the wavefunction has six independent
components, i.e. there is doubling of the states of a vector particle with
spin projections $s_z=\pm 1,0$.

We proceed from the two (for $\varepsilon =\pm 1$) wave equations for
arbitrary spin particles with the EDM and AMM in external electromagnetic
fields:

\begin{equation}
\left[ \mathcal{D}_\mu ^2-m^2-\frac e2\left( gF_{\mu \nu }-\sigma \widetilde{
F}_{\mu \nu }\right) \Sigma _{\mu \nu }^{(\varepsilon )}\right] \Psi
_\varepsilon (x)=0,  \label{1}
\end{equation}

where $\mathcal{D}_\mu =\partial _\mu -ieA_\mu $ is the covariant
derivative, $\partial _\mu =\partial /\partial x_\mu $, $x_\mu =(\mathbf{x}
,ix_0)$ ($x_0$ is the time, $x_0\equiv t$), $F_{\mu \nu }=\partial _\mu
A_\nu -\partial _{_\nu }A_\mu $ is the electromagnetic field strength
tensor, $\widetilde{F}_{\mu \nu }=(1/2)\epsilon _{\mu \nu \alpha \beta
}F_{\alpha \beta }$ is the dual tensor ($\epsilon _{1234}=-i$), $\epsilon
_{\mu \nu \alpha \beta }$ is the Levi-Civita symbol. The numbers $%
\varepsilon =\pm 1$ correspond to the $\left( s,0\right) $ and $\left(
0,s\right) $ representations with the generators of the Lorenz group $\Sigma
_{\mu \nu }^{\left( -\right) }$, $\Sigma _{\mu \nu }^{\left( +\right) }$,
respectively. The gyromagnetic ratio $g=1/s+\kappa $ ($\kappa $ is the AMM
of a particle) and the $\sigma $ give the contribution to the magnetic
moment $\mu =egs/(2m)$ and the EDM of arbitrary spin particles, $d=\sigma
/(2m)$. The spin matrices $S_k$ are connected with the generators $%
\Sigma_{\mu \nu }^{(\varepsilon )}$ by the relationships:

\begin{equation}
\Sigma _{ij}^{(\varepsilon )}=\epsilon _{ijk}S_k,\hspace{0.5in}\Sigma
_{4k}^{(\varepsilon )}=-i\varepsilon S_k  \label{2}
\end{equation}

and obey the commutation relations

\begin{equation}
\left[ S_i,S_j\right] =i\epsilon _{ijk}S_k,\hspace{0.5in}\left( S_1\right)
^2+\left( S_2\right) ^2+\left( S_3\right) ^2=s(s+1)  \label{3}
\end{equation}

with $i,j,k=1,2,3$ ($\epsilon _{123}=1$).

Eq. (1) at the case $\sigma =0$ was investigated in [18], and at
the particular case of the ``normal'' magnetic moment of a
particle, when $g=1/s$ , we arrive at the approach [10]. Here
arbitrary AMM and EDM of a particle is considered. At the parity
transformation, $\varepsilon \rightarrow -\varepsilon ,$ and the
representation $(s,0)$ is converted into $(0,s)$. As a result Eqs.
(1) (at $\varepsilon =\pm 1$) are invariant under the parity
inversion if $\sigma =0$. The term with the EDM in Eq. (1)
violates the $CP$ - invariance.

Now we find the solutions to Eq. (1) for a particle in the field
of uniform and constant electromagnetic fields. For simplicity we
choose a coordinate system in which the electric $\mathbf{E}$ and
magnetic $\mathbf{H}$ fields are parallel, so that
$\mathbf{E}=\mathbf{n}E,$ $\mathbf{H}=\mathbf{n}H$, $
\mathbf{n}=(0,0,1)$ with the 4-vector potential

\begin{equation}
A_\mu =\left( 0,x_1H,-tE,0\right).  \label{4}
\end{equation}

The Lorentz condition $\partial_\mu A_\mu =0$ is permitted for the
potential (4). Different choices of potentials for the fixed
electromagnetic fields $ \mathbf{E}$, $\mathbf{H}$ are connected
by the gauge transformations. When $ E\neq H$, two Lorentz
invariants of electromagnetic fields

\begin{equation}
\mathcal{F}=\frac 14F_{\mu \nu }^2=\frac 12\left( \mathbf{H}^2-\mathbf{E}
^2\right) ,\hspace{0.3in}\mathcal{G}=\frac 14F_{\mu \nu }\widetilde{F}_{\mu
\nu }=\mathbf{E}\cdot \mathbf{H},  \label{5}
\end{equation}

do not vanish. This is the general case of the external uniform and constant
electromagnetic fields. From Eqs. (2) we arrive at the expressions
containing spin matrices and auxiliary complex vector fields:

\[
\frac 12\Sigma _{\mu \nu }^{(+)}F_{\mu \nu }=\mathbf{SX},\hspace{1.0in}\frac
12\Sigma _{\mu \nu }^{(-)}F_{\mu \nu }=\mathbf{SX}^{*},
\]
\vspace{-8mm}
\begin{equation}
\label{6}
\end{equation}
\vspace{-8mm}
\[
\frac 12\Sigma _{\mu \nu }^{(+)}\widetilde{F}_{\mu \nu
}=\mathbf{S} \widetilde{\mathbf{X}},\hspace{1.0in}\frac 12\Sigma
_{\mu \nu }^{(-)} \widetilde{F}_{\mu \nu
}=\mathbf{S}\widetilde{\mathbf{X}}^{*},
\]
where $\mathbf{X}=\mathbf{H}+i\mathbf{E}$, $\mathbf{X}^{*}=\mathbf{H}-i
\mathbf{E}$, $\widetilde{\mathbf{X}}=-\mathbf{E}+i\mathbf{H}$, $\widetilde{
\mathbf{X}}^{*}=-\mathbf{E}-i\mathbf{H}$. The auxiliary complex vector
fields $\mathbf{X}$, $\widetilde{\mathbf{X}}$ (and $\mathbf{X}^{*}$,
$\widetilde{\mathbf{X}}^{*}$) are transformed as $\mathbf{X}\rightarrow
\widetilde{\mathbf{X}}$ under the dual transformations of the
electromagnetic fields $\mathbf{E}\rightarrow \mathbf{H}$, $\mathbf{H}
\rightarrow -\mathbf{E}$.

Let the wavefunction $\Psi _\varepsilon (x)$ be the eigenfunction of the
operators $\mathbf{SX}$ and $\mathbf{S}\widetilde{\mathbf{X}}$. Then we have
the diagonal representation of matrices (6), and equations for eigenvalues
are given by

\[
\mathbf{SX}\Psi _{+}^{(s_z)}(x)=s_zX\Psi _{+}^{(s_z)}(x),\hspace{0.5in}
\mathbf{SX}^{*}\Psi _{-}^{(s_z)}(x)=s_zX^{*}\Psi _{-}^{(s_z)}(x),
\]
\vspace{-8mm}
\begin{equation}  \label{7}
\end{equation}
\vspace{-8mm}
\[
\mathbf{S}\widetilde{\mathbf{X}}\Psi _{+}^{(s_z)}(x)=s_z\widetilde{X}\Psi
_{+}^{(s_z)}(x),\hspace{0.5in}\mathbf{S}\widetilde{\mathbf{X}}^{*}\Psi
_{-}^{(s_z)}(x)=s_z\widetilde{X}^{*}\Psi _{-}^{(s_z)}(x),
\]

where $X=H+iE$, $\widetilde{X}=-E+iH$, and the spin projections $s_z$ take
the values

\begin{equation}
s_z=\biggl \{
\begin{array}{c}
\pm s,\pm (s-1),\cdot \cdot \cdot 0 \\
\pm s,\pm (s-1),\cdot \cdot \cdot \pm \frac 12
\end{array}
\hspace{0.5in}
\begin{array}{c}
\mbox{{for~bosons}}, \\
\mbox{{for~fermions}}.
\end{array}
\label{8}
\end{equation}

Using Eqs. (6), (7), we can represent Eq. (1) (for $\varepsilon =\pm 1$) as

\[
\left[ \mathcal{D}_\mu ^2-m^2-es_z\left( gX-\sigma \widetilde{X}\right)
\right] \Psi _{+}^{(s_z)}(x)=0,
\]
\vspace{-8mm}
\begin{equation}  \label{9}
\end{equation}
\vspace{-8mm}
\[
\left[ \mathcal{D}_\mu ^2-m^2-es_z\left( gX^{*}-\sigma \widetilde{X}
^{*}\right) \right] \Psi _{-}^{(s_z)}(x)=0,
\]

where $s_z$ is given by Eq. (8). For every projection $s_z$ equations (9)
are the Klein-Gordon type equations with the complex ``effective'' masses:
\[
m_{eff}^2=m^2+es_z\left( gX-\sigma \widetilde{X}\right),
\]
\vspace{-8mm}
\begin{equation}  \label{10}
\end{equation}
\vspace{-8mm}
\[
(m_{eff}^2)^{*}=m^2+es_z\left( gX^{*}-\sigma \widetilde{X}^{*}\right).
\]

Introducing the variables [19] (see also [20])

\[
\eta =\frac{p_2-eHx_1}{\sqrt{eH}},\hspace{0.5in}\tau =\sqrt{eE}\left( t+
\frac{p_3}{eE}\right),
\]
\vspace{-8mm}
\begin{equation}  \label{11}
\end{equation}
\vspace{-8mm}
\[
\Psi _{+}^{(s_z)}(x)\equiv \Psi ^{(s_z)}(x)=\exp \left[ i\left(
p_2x_2+p_3x_3\right) \right] \Phi ^{(s_z)}(\eta ,\tau ),
\]

Eq. (9) becomes

\begin{equation}
\left[ eH\left( \partial _\eta ^2-\eta ^2\right) -eE\left( \partial _\tau
^2+\tau ^2\right) -m^2-es_z\left( gX-\sigma \widetilde{X}\right) \right]
\Phi ^{(s_z)}(\eta ,\tau )=0,  \label{12}
\end{equation}

plus complex conjugated equation where $\partial _\eta =\partial /\partial
\eta $, $\partial _\tau =\partial /\partial \tau $. The variables $\eta $,
$\tau $ are separated in this equation and the solution to Eq. (12) can be
written in the form

\begin{equation}
\Phi ^{(s_z)}(\eta ,\tau )=\phi ^{(s_z)}(\eta )\chi ^{(s_z)}(\tau ),
\label{13}
\end{equation}

with the eigenfunctions $\phi ^{(\lambda )}(\eta )$, $\chi (\tau )$ obeying
the following equations

\begin{equation}
\left[ eH\left( \partial _\eta ^2-\eta ^2\right) -m^2-es_z\left( gX-\sigma
\widetilde{X}\right) +k_{s_z}^2\right] \phi ^{(s_z)}(\eta )=0,  \label{14}
\end{equation}
\begin{equation}
\left[ eE\left( \partial _\tau ^2+\tau ^2\right) +k_{s_z}^2\right] \chi
^{(s_z)}(\tau )=0,  \label{15}
\end{equation}

It follows from Eqs. (14), (15) that a magnetic field $H$ forms a motion of
a particle depending on the variables $x_1$, $x_2$ and an electric field
determines a motion depending on $x_3$, $t$. In accordance with Eqs.
(13)-(15) these two motions are described by the functions $\phi
^{(s_z)}(\eta )$, $\chi ^{(s_z)}(\tau )$ and they are independent. The
solution to Eq. (14) is given by the Hermite functions [21]

\begin{equation}
\phi ^{(s_z)}(\eta )=N_0\exp \left( -\frac{\eta ^2}2\right) H_n(\eta )),
\label{16}
\end{equation}

which is finite at $\eta \rightarrow \infty $. Here $N_0$ is the
normalization constant, and $H_n(\eta )$ are the Hermite polynomials. The
requirement that the solution to Eq. (14) be finite leads to the equation
for eigenvalues $k_{s_z}^2$:

\begin{equation}
k_{s_z}^2-m^2-es_z\left( gX-\sigma \widetilde{X}\right) =eH(2n+1),
\label{17}
\end{equation}

where $n=1,2,...$ is the principal quantum number. So, the spectral
parameter $k_{s_z}^2$ is a quantized but not an arbitrary value.

Solutions to Eq. (15) can be expressed through the parabolic-cylinder
functions $D_\nu (x)$ [21] as

\[
_{+}\chi ^{(s_z)}(\tau )=D_\nu [-(1-i)\tau ],\hspace{0.3in}^{-}\chi
^{(s_z)}(\tau )=D_\nu [(1-i)\tau ],
\]
\vspace{-8mm}
\begin{equation}  \label{18}
\end{equation}
\vspace{-8mm}
\[
^{+}\chi ^{(s_z)}(\tau )=D_{-(\nu+1)}[(1+i)\tau ],\hspace{0.3in}_{-}\chi
^{(s_z)}(\tau )=D_{-(\nu+1)}[-(1+i)\tau ],
\]

Four solutions (18) possess different asymptotic at $t\rightarrow \pm \infty
$ [19], and $\nu $ reads

\[
\nu =\frac{ik_{s_z}^2}{2eE}-\frac 12.
\]

Using Eqs. (13), (16) and (18), we arrive at the solutions of Eqs. (9):

\begin{equation}
_{\pm }^{\pm }\Psi ^{(s_z)}(x)=N_0\exp \left\{ i(p_2x_2+p_3x_3)-\frac{\eta
^2 }2\right\} H_n(\eta )_{\pm }^{\pm }\chi ^{(s_z)}(\tau ).  \label{19}
\end{equation}

The solutions $_{\pm }^{\pm }\Psi ^{(s_z)}(x)$ are labelled by four
conserved numbers: two momentum projections $p_2$, $p_3$, the spin
projection $s_z$ and the spectral parameter $k_{s_z}$. Wavefunctions (19)
are the exact solutions of Eq. (9) for a particle in external constant and
uniform electromagnetic fields corresponding to the principal quantum number
$n$ and the spin projection $s_z$. The exact solutions (19) can be used for
different electrodynamic calculations of the quantum processes with the
presence of arbitrary spin particles. In such approach the interaction with
an external electromagnetic field is taken into consideration with its exact
value, and therefore the calculations may be used without the perturbative
theory in a small parameter. It allows us to investigate some
nonperturbative and nonlinear effects.

\section{Pair production of arbitrary spin particles by
electromagnetic fields}

According to the well known Schwinger result [22] a uniform and constant
electric field produces pairs of particles. The probabilities for the
production of scalar and spin 1/2 pairs of particles have been evaluated.
Now we consider more general case of pair production of arbitrary spin
particles with the EDM and AMM by electric and magnetic fields. The
probability of pair production of particles by constant and uniform
electromagnetic fields may be obtained using the asymptotic form of the
solutions [19]. In this approach one avoids the Klein ``paradox" when the
total scattering probability of a wave packet of a particle is smaller than
unity.

The functions $_{+}^{+}\chi ^{(s_z)}(\tau )$ (19) have positive frequency
and $_{-}^{-}\chi ^{(s_z)}(\tau )$ have negative frequency when the time
approaches $\pm \infty $. The set of wavefunctions $^{\pm}\chi ^{(s_z)}(\tau
)$ and $_{\pm}\chi ^{(s_z)}(\tau )$ (19) are equivalent and satisfy the
relationships [19]:

\[
_{+}\Psi ^{(s_z)}(x)=c_{1ns_z}{}^{+}\Psi ^{(s_z)}(x)+c_{2ns_z}{}^{-}\Psi
^{(s_x)}(x),
\]
\[
^{+}\Psi ^{(s_z)}(x)=c_{1ns_z}^{*}{}_{+}\Psi ^{(s_z)}(x)-c_{2ns_z}{}_{-}\Psi
^{(s_z)}(x),
\]
\vspace{-8mm}
\begin{equation}  \label{20}
\end{equation}
\vspace{-8mm}
\[
_{-}\Psi ^{(s_z)}(x)=c_{2ns_z}^{*}{}^{+}\Psi
^{(s_z)}(x)+c_{1ns_z}^{*}{}^{-}\Psi ^{(s_z)}(x),
\]
\[
^{-}\Psi ^{(s_z)}(x)=-c_{2ns_z}^{*}{}_{+}\Psi
^{(s_z)}(x)+c_{1ns_z}{}_{-}\Psi ^{(s_z)}(x),
\]

where variables $c_{1ns_z},$ $c_{2ns_z}$ are

\[
c_{2ns_z}=\exp \left[ -\frac \pi 2(\lambda +i)\right] ,\hspace{0.3in}\lambda
=\frac{m^2+es_z\left( gX-\sigma \widetilde{X}\right) +eH(2n+1)}{eE},
\]
\begin{equation}
\mid c_{1ns_z}\mid ^2-\mid c_{2ns_z}\mid
^2=1\hspace{1.0in}\mbox{for~bosons},  \label{21}
\end{equation}
\[
\mid c_{1ns_z}\mid ^2+\mid c_{2ns_z}\mid ^2=1\hspace{1.0in}\mbox{
for~fermions}.
\]

The relations (20) represent the Bogolubov transformations with
the coefficients $c_{1ns_z}$, $c_{2ns_z}$ which contain the
information about producing pairs of particles in the state with
the principal quantum number $ n$ and the spin projection $s_z$.
So, the value $\mid c_{2ns_z}\mid ^2$ is the probability of pair
production of arbitrary spin particles in the state with quantum
numbers $n$, $s_z$ throughout all space and during all time, and
according to Eq. (21) is given by

\begin{equation}
\mid c_{2ns_z}\mid ^2=\exp \left\{ -\pi \left[ \frac{m^2}{eE}+s_z\left(
\sigma +g\frac HE\right) +\frac HE(2n+1)\right] \right\}.  \label{22}
\end{equation}

The expression (22) represents also the probability of the pair annihilation
with quantum numbers $n,$ $s_z$. It follows from Eq. (22) that the maximum
of pair creation probability at $H\gg E$ occurs in the state with the
smallest energy when $n=0,$ $s_z=-s$. According to the approach [19], the
average number of particle pairs produced from a vacuum is given by

\begin{equation}
\overline{N}=\sum_{n,s_z}\mid c_{2ns_z}\mid ^2\frac{e^2EHVT}{(2\pi )^2},
\label{23}
\end{equation}

where $V$ is the normalization volume, $T$ is the time of observation.

Inserting expression (22) into Eq. (23) and calculating the sum over the
principal quantum number $n$ and spin projections $s_z$ (see [18]), we find
the pair production probability per unit volume and per unit time

\begin{equation}
I(E,H)=\frac{\overline{N}}{VT}=\frac{e^2EH}{8\pi ^2}\frac{\exp \left[ -\pi
m^2/(eE)\right] }{\sinh \left( \pi H/E\right) }\frac{\sinh \left[ (s+1/2)\pi
\left( \sigma +gH/E\right) \right] }{\sinh \left[ \left( \pi /2\right)
\left( \sigma +gH/E\right) \right] }.  \label{24}
\end{equation}

The expression (24) is non-analytic in $E$ and it is impossible to
obtain it with the help of a perturbative theory. We find from Eq.
(24) that in the case $\sigma =g=0$, the pair production of
arbitrary spin particles is ($ 2s+1)$ times that for the scalar
particle pair production due to the ($2s+1)$ physical degrees of
freedom of the arbitrary spin field. At $\sigma =0$ Eq. (24)
converts into the expression derived in [18]. The intensity of the
pair creation (24) is the generalization of the results [11,18] on
the case of arbitrary spin particles with the EDM and AMM. In the
general case $\sigma \neq 0$, $g\neq 0$ there is also a pair
production of arbitrary spin particles if $E=0$, $H>m^2/e$ at
$gs>1$ (see [11]), i.e. there is instability of the vacuum in a
magnetic field. This occurs because the ground state with quantum
numbers $n=0$, $s_z=-s$ becomes a tachyon state when the magnetic
field $H$ is greater than the critical value $H_0=m^2/e$, i.e.
$H>H_0$, and the vacuum is reconstructed.

When the magnetic field vanishes, $H=0$, equation (24) transforms into

\begin{equation}
I(E)=\frac{e^2 E^2}{8\pi ^3} \frac{\exp \left( -\pi E_0/E \right)} {\sinh
\left[\pi \sigma/2 \right] }\sinh \left[ (2s+1)\pi \sigma /2 \right],
\label{25}
\end{equation}

where $E_0=m^2 /e$ is the critical value of the electric field. At
$E<E_0$ the intensity of pair production is exponentially small.
However at the critical value of the electric field $E=E_0$
particles are created rapidly. So, the intensity for the
electron-positron pair production ($s=1/2$, $ \sigma =0$) at
$E=m_e^2 /e$, where $m_e$ is the mass of the electron, is given by

\[
I(E_0)=\frac{m_e^4}{4\pi^3}\exp(-\pi)\simeq4\times 10^{48}~cm^{-3}sec^{-1}.
\]

The pair production probability (25) does not depend on the gyromagnetic
ratio $g$ and depends only on the EDM. It follows from Eq. (24) that the
magnetic field increases the producing of higher spin particles and
decreases the pair production of scalar particles.

Let us consider the important cases of particles with spins $1/2$ and $1$.
For fermions with spin $1/2$ the pair production probability by the
electromagnetic field takes the form

\begin{equation}
I_{1/2}(E,H)=\frac{e^2EH}{4\pi ^2}\frac{\exp \left[ -\pi
m^2/(eE)\right] }{ \sinh \left( \pi H/E\right) }\cosh \left[
\left( \pi /2\right) \left( \sigma +gH/E\right) \right].
\label{26}
\end{equation}

For the probability of electron-positron production we have $g=2$,
$\sigma=0$ and Eq. (26) converts into those obtained in [19]. At
the particular case of an external electric field we arrive at the
Schwinger formula (for $g=2$, $ \sigma=0$) [22]. The presence of
the EDM and AMM of a spinor particle increases the pair
production. From the general expression (24) we find after some
transformations the intensity of producing of the vector particles
($s=1$)

\begin{equation}
I_{1}(E,H)=\frac{e^2EH}{8\pi ^2}\frac{\exp \left[ -\pi
m^2/(eE)\right] }{ \sinh \left( \pi H/E\right) }\left[2\cosh
\left[ \pi \left( \sigma +gH/E\right)\right]+1\right].  \label{27}
\end{equation}

The pair production probability (27) is greater than those for spinor
particles (26) due to the greater number of physical degrees of freedom
(spin projections) of the vector field. At $\sigma=0$ and $H=0$ expression
(27) converts into the intensity of pair production of vector particles
obtained in [14] for the case $g=2$.

The imaginary part of the density of the electromagnetic field Lagrangian
can be obtained with the help of [19]:

\begin{equation}
\mbox{Im}\mathcal{L}=\frac 12\int \sum_{n,s_z}\ln \mid
c_{1ns_z}\mid ^2\frac{ e^2EH}{(2\pi )^2}.  \label{28}
\end{equation}

Using Eqs. (21) we find

\[
\mbox{Im}\mathcal{L}=\frac{e^2EH}{16\pi ^2}\sum_{n=1}^\infty
\frac{\beta _n} n\exp \left( -\frac{\pi m^2n}{eE}\right)
\frac{\sinh \left[ n(s+1/2)\pi \left( \sigma +gH/E\right) \right]
}{\sinh \left( n\pi H/E\right) \sinh \left[ n\left( \pi /2\right)
\left( \sigma +gH/E\right) \right] },
\]
\vspace{-8mm}
\begin{equation}  \label{29}
\end{equation}
\vspace{-8mm}
\[
\beta _n=\biggl \{
\begin{array}{c}
(-1)^{n-1}\hspace{1.0in}\mbox{{for~integer spins}}, \\
1\hspace{1.0in}\mbox{{for~half-integer spins}}.
\end{array}
\]

In accordance with [22,19] the first term ($n=1$) in (29) gives the half of
the pair production probability per unit volume and per unit time. It should
be noted that Im$\mathcal{L}$ (29) and the pair production probabilities
(24)-(27) do not depend on the scheme of renormalization as all divergences
and the renormalizability are contained in Re$\mathcal{L}$ [22].

\section{Polarization of arbitrary spin particle vacuum}

Let us consider the problem of one-loop corrections to the Lagrange function
of a constant and uniform electromagnetic field due to the field interaction
with a vacuum of arbitrary spin particles with the EDM and AMM. This problem
has been solved for fields of spins $0$, $1/2$ and $1$ (at $\sigma =0$, $g=2$
) in [22,14,15]. The nonlinear corrections to the Maxwell Lagrangian due to
vacuum polarization are connected with the cross-section of scattering
photons by photons. Using the Schwinger method [22], we obtain the one-loop
corrections to Lagrangian of a constant and uniform electromagnetic field
\footnote{The factor $\epsilon $ was omitted in [18].}

\begin{equation}
\mathcal{L}^{(1)}=\frac \epsilon {32\pi ^2}\int_0^\infty d\tau
\tau ^{-3}\exp \left( -m^2\tau -l(\tau )\right) \mbox{tr}\exp
\left[ \frac{ e_0\tau }2\Sigma _{\mu \nu }\left( gF_{\mu \nu
}-\sigma \widetilde{F}_{\mu \nu }\right) \right],  \label{30}
\end{equation}
\[
\Sigma _{\mu \nu }=\Sigma _{\mu \nu }^{(+)}\oplus \Sigma _{\mu \nu
}^{(-)}, \hspace{0.3in}l(\tau )=\frac 12\mbox{tr}\ln \left[ \left(
e_0F\tau \right) ^{-1}\sin (e_0F\tau )\right],
\]
\vspace{-8mm}
\begin{equation}  \label{31}
\end{equation}
\vspace{-8mm}
\[
\exp \left[ -l(\tau )\right] =\frac{(e_0\tau
)^2\mathcal{G}_0}{\mbox{Im} \cosh (e_0\tau X_0)},
\]

where $\oplus $ is the direct sum, $\epsilon =1$ for bosons and
$\epsilon =-1 $ for fermions due to different signs of loop
integrals for bosons and fermions. The index $0$ in Eqs. (30),
(31) refers to the unrenormalized variables, so that $e_0$ is the
bare electric charge, $\mathbf{X}_0=\mathbf{ H }_0+i\mathbf{E}_0$,
$\mathcal{G}_0=\mathbf{E}_0\cdot \mathbf{H}_0$; $ \mathbf{\ E}_0$,
$\mathbf{H}_0$ are bare electric and magnetic fields,
respectively. The Lagrangian $\mathcal{L}^{(1)}$ (30) is the
effective nonlinear Lagrangian which is an integral over the
proper time $\tau $. Calculating with the help of Eqs. (6), (7)
the trace (tr) of the matrices we arrive at

\begin{equation}
\mbox{tr}\exp \left[ \frac{e_0\tau }2\Sigma _{\mu \nu }\left(
gF_{\mu \nu }-\sigma \widetilde{F}_{\mu \nu }\right) \right]
=2\mbox{Re}\frac{\sinh \left[ (s+1/2)e_0\tau \left( gX_0-\sigma
\widetilde{X}_0\right) \right] } {\sinh \left[ \left( e_0\tau
/2\right) \left( gX_0-\sigma \widetilde{X} _0\right) \right] }.
\label{32}
\end{equation}

Replacing (32) into (30) and subtracting the constant to have
vanishing $ \mathcal{L}^{(1)}$ when the electromagnetic fields
approach zero, (see [22]) we arrive at

\[
\mathcal{L}^{(1)}=\frac \epsilon {16\pi ^2}\int_0^\infty d\tau \tau
^{-3}\exp \left( -m^2\tau \right)
\]
\vspace{-8mm}
\begin{equation}  \label{33}
\end{equation}
\vspace{-8mm}
\[
\times \left\{ \frac{(e_0\tau )^2\mathcal{G}_0}{\mbox{Im}\cosh
(e_0\tau X_0)} \mbox{Re}\frac{\sinh \left[ (s+1/2)e_0\tau \left(
gX_0-\sigma \widetilde{X} _0\right) \right] }{\sinh \left[ \left(
e_0\tau /2\right) \left( gX_0-\sigma \widetilde{X}_0\right)
\right] }-\left( 2s+1\right) \right\}.
\]

Setting $\sigma =0$, $g=2$, $\epsilon =-1$ in Eq. (33) we come to the
Schwinger Lagrangian [22]. The integral (30) is the nonlinear correction to
Maxwell's Lagrangian due to the vacuum polarization of arbitrary spin
particles with the EDM and AMM. The Lagrangian (33) has the quadratic term
in the electromagnetic fields, and that renormalizes the Lagrangian of the
free electromagnetic fields
\begin{equation}
\mathcal{L}^{(0)}=-\mathcal{F}_0\mathcal{=}\frac 12\left( \mathbf{E}_0^2-
\mathbf{H}_0^2\right).  \label{34}
\end{equation}

Expanding Lagrangian (33) in weak electromagnetic fields and adding the
Lagrangian of the free electromagnetic fields (34) we obtain the
renormalized Lagrangian of electromagnetic fields that takes into account
the vacuum polarization of arbitrary spin particles with the EDM and AMM:

\[
\mathcal{L}=\mathcal{L}^{(0)}+\mathcal{L}^{(1)}=-\mathcal{F}+\frac \epsilon
{16\pi ^2}\int_0^\infty d\tau \tau ^{-3}\exp \left( -m^2\tau \right)
\]
\begin{equation}
\times \biggl \{ \frac{(e\tau )^2\mathcal{G}}{\mbox{Im}\cosh
(e\tau X)}\mbox{ Re}\frac{\sinh \left[ (s+1/2)e\tau \left(
gX-\sigma \widetilde{X}\right) \right] }{\sinh \left[ \left( e\tau
/2\right) \left( gX-\sigma \widetilde{X} \right) \right] }
\label{35}
\end{equation}
\[
-\left( 2s+1\right) -\frac{\left( 2s+1\right) (e\tau )^2\mathcal{F}}3\left[
s(s+1)\left( g^2-\sigma ^2\right) -1\right] \biggl \},
\]
where we renormalize fields and charges:
\[
\mathcal{F}=Z_3^{-1}\mathcal{F}_0,\hspace{0.3in}e=Z_3^{1/2}e_0.
\]

The renormalization constant is given by

\begin{equation}
Z_3^{-1}=1-\frac{\epsilon e_0^2\left( 2s+1\right) \left[ s(s+1)\left(
g^2-\sigma ^2\right) -1\right] }{48\pi ^2}\int_0^\infty d\tau \tau ^{-1}\exp
\left( -m^2\tau \right).  \label{36}
\end{equation}

If the electromagnetic fields $\mathbf{E}$ , $\mathbf{H}$ are absent,
Lagrangian (35) vanishes. The renormalization constant $Z_3^{-1}$ diverges
logarithmically when the cutoff factor $\tau _0$ at the lower limit in the
integral (36) approaches zero ($\tau _0\rightarrow 0$). It follows from Eq.
(36) that if the inequality

\begin{equation}
\epsilon \left[ s(s+1)\left( g^2-\sigma ^2\right) -1\right] >0  \label{37}
\end{equation}

is valid, the renormalization constant of the charge $Z_3^{1/2}>1$. This
case indicates asymptotic freedom in the field [23,24], and is not realized
in QED because $g=2$, $\sigma =0$ and $\epsilon =-1$. However, for boson
fields, when $\epsilon =1$, asymptotic freedom occurs for the small value of
the EDM $\sigma $. In accordance with Eq. (37) the asymptotically free
behavior in the boson fields is due to the AMM, and the EDM of particles
suppresses the phenomena of asymptotic freedom. However this discussion
concerns only the renormalizable theories. We can argue that the formal
counting of the divergences corresponding to Eq. (1) leads to a
renormalizable theory due to the form of the field propagator that
proportional to $1/p^2$, and the smallness of the field self-interaction
constant. If $g=2$, $\sigma =0$, we have the linear approximation to the
renormalizable gage theory for the vector field when the mass of the field
is acquired by the Higgs mechanism. Eq. (36) allows us to obtain the
Callan-Zymanzik $\beta $ function that corresponds to the renormalizable
theory:

\begin{equation}
\beta =-\frac{\epsilon e_0^2\left( 2s+1\right) \left[ s(s+1)\left(
g^2-\sigma ^2\right) -1\right] }{48\pi ^2}.  \label{38}
\end{equation}

Asymptotic freedom occurs if the $\beta $ function is negative ($\beta <0$)
that is equivalent to Eq. (37). The AMM and spin ($s$) of bosons assures
asymptotic freedom and instability of the vacuum in a magnetic field.

Expanding Eq. (35) in the weak electromagnetic fields, and using the equality

\[
\int_0^\infty d\tau \tau \exp \left( -m^2\tau \right) =\frac 1{m^4}
\]
we obtain after renormalization the Maxwell Lagrangian with the nonlinear
corrections:
\[
\mathcal{L}=\frac 12\left( \mathbf{E}^2-\mathbf{H}^2\right) -\frac{
2s(s+1)\sigma g}{s(s+1)(g^2-\sigma ^2)-1}\left( \mathcal{G}-\mathcal{G}
_0\right) +\frac{\epsilon \alpha ^2(2s+1)}{90m^4}
\]
\[
\times \biggl \{
\left[ s(s+1)(3s^2+3s-1)\left( g^4-6g^2\sigma ^2+\sigma ^4\right)
-10s(s+1)(g^2-\sigma ^2)+7\right] \mathcal{F}^2
\]
\vspace{-8mm}
\begin{equation}  \label{39}
\end{equation}
\vspace{-8mm}
\[
+\left[ s(s+1)(3s^2+3s-1)\left( 6g^2\sigma ^2-g^4-\sigma ^4\right) +1\right]
\mathcal{G}^2
\]
\[
+4s(s+1)\sigma g\left[ 2(3s^2+3s-1)\left( g^2-\sigma ^2\right) -5\right]
\mathcal{GF}\biggr \},
\]

where $\alpha =e^2/(4\pi )$. The second and last terms in Eq. (39) indicate
parity violation due to the EDM of a particle. We can consider the second
term in Eq. (39) as anomaly for a particle with the EDM because such a
quadratic in fields term does not present in the bare Lagrangian (34).
Lagrangian (39) is the Heisenberg-Euler type Lagrangian [25,26] for the case
of arbitrary spin particles with the EDM and AMM. It can be verified that
for the case $\sigma =g=0$, we arrive from Eq. (39) at

\begin{equation}
\mathcal{L}(\sigma =g=0)=\frac 12\left(
\mathbf{E}^2-\mathbf{H}^2\right) +\epsilon
(2s+1)\mathcal{L}_{\mbox{spin 0}},  \label{40}
\end{equation}

where $\mathcal{L}_{\mbox{spin 0}}$ is the nonlinear correction to
the Lagrangian of the electromagnetic fields due to the vacuum
polarization of scalar particles [22]:

\begin{equation}
\mathcal{L}_{\mbox{spin 0}}=\frac{\alpha ^2}{360m^4}\left[ 7\left(
\mathbf{E} ^2-\mathbf{H}^2\right) ^2+4(\mathbf{EH})^2\right]
\label{41}
\end{equation}

Eq. (40) tells us that there is an equal contribution of $(2s+1)$ degrees of
freedom (spin projections) of arbitrary spin fields at $\sigma =g=0$ when
Eq. (1) is converted into a Klein-Gordon equation. The factor $\epsilon$
corresponds to different statistics for bosons and fermions. For the case of
QED at $s=1/2,$ $g=2$, $\epsilon =-1$ Eq. (39) becomes the Schwinger
Lagrangian [22].

It is interesting to consider the particular case of spin $1$ fields with
gyromagnetic ratio $g=2$ and $\sigma =0$ which corresponds to the linear
approximation to the renormalizable SM. However, it should be noted that Eq.
(1) at $s=1$ is based on the $\left( 1,0\right) \oplus \left( 0,1\right)$
representation of the Lorenz group, and is not equivalent to the Proca
equation [18]. Setting $g=2$, $\sigma =0$, $s=1$, $\epsilon =1$ in Eq. (39)
we arrive at the Lagrangian of the electromagnetic field that takes into
account the vacuum polarization of a charged vector particles

\begin{equation}
\mathcal{L}_{\mbox{spin 1}}=\frac 12\left(
\mathbf{E}^2-\mathbf{H}^2\right)+ \frac{\alpha ^2}{10m^4}\left[
\frac{29}4\left( \mathbf{E}^2-\mathbf{H} ^2\right)
^2-53(\mathbf{EH})^2\right]  \label{42}
\end{equation}

which is a little different from those obtained in [14] on the basis of the
Proca equation.

Eq. (35) allows us to find the limit at $eE/m^2\rightarrow \infty $ and
$eH/m^2\rightarrow \infty $ if we ignore the dependence of the AMM and EDM on
the strong external electromagnetic fields that is some approximation.

\section{Conclusion}

The pair-production probability (24), and the effective Lagrangian for
electromagnetic fields (35) which takes into account the polarization of the
vacuum, are the generalization of the Schwinger result on the case of the
theory of particles with arbitrary spins, EDM and AMM in the external
electric and magnetic fields. It follows from Eq. (24) that there is a pair
production of particles by a purely magnetic field ($H>H_0$) in the case of
$gs>1$ assuring asymptotic freedom and instability of the vacuum in a
magnetic field. The presence of the EDM of a particle does not lead to
instability of the vacuum in a magnetic field but it suppresses the
phenomena of asymptotic freedom. In the presence of the magnetic field the
probability decreases for scalar particles and increases for higher spin
particles.

The intensity of pair production of arbitrary spin particles (24) does not
depend on the renormalization scheme as all divergences and the
renormalizability are contained in Re$\mathcal{L}$. Therefore the formula
(24) is reliable. The procedure of obtaining the vacuum polarization
corrections for electromagnetic fields uses the renormalization, and we
imply that the scheme considered is the linearized version of renormalizable
gauge theory. This point of view is justified if we imply the smallness of
the arbitrary spin field self-interaction constant.

We have just studied the vacuum quantum effects of pair production and
vacuum polarization of arbitrary spin particles with the EDM which violates
$CP$ - symmetry. The investigation of quantum processes with $CP$ - violation
is important because they may give a sensitive probe for New Physics. The
value of the gyromagnetic ratio $g\neq 2$ for vector particles, and the EDM
$d \neq 0$ leads to physics beyond the SM. Now this is of interest because
experimental muon AMM data [27] have challenged the SM (there is a
discrepancy of $2.6\sigma $ deviation between the theory and the averaged
experimental value).

\end{document}